\documentclass[10pt,a4paper]{book} 

\usepackage{helvet,natbib,color,lastpage}
\setcitestyle{aysep={}} 

\usepackage{longtable}
\usepackage{fancyhdr}
\usepackage{xspace,enumitem}
\usepackage{multicol}
\usepackage{makecell}
\usepackage{anyfontsize}
\usepackage{epsfig,caption}
\usepackage{bibentry}
\usepackage{tcolorbox}

\usepackage{graphicx}
\usepackage{lipsum}  
\usepackage[default]{lato}
\usepackage[T1]{fontenc}
\usepackage[letterspace=50]{microtype}
\usepackage[official]{eurosym}
\usepackage{datetime}

\usepackage[breaklinks,colorlinks,citecolor=blue]{hyperref}
\usepackage[all]{hypcap}

\usepackage{array}
\newcommand{\PreserveBackslash}[1]{\let\temp=\\#1\let\\=\temp}
\newcolumntype{C}[1]{>{\PreserveBackslash\centering}p{#1}}
\newcolumntype{R}[1]{>{\PreserveBackslash\raggedleft}p{#1}}
\newcolumntype{L}[1]{>{\PreserveBackslash\raggedright}p{#1}}

\newdateformat{monthyeardate}{\monthname[\THEMONTH] \THEYEAR}

\newcommand{\cool}{COOL Research DAO\xspace}

\newcommand{\fsize}{10}
\newcommand{\skipsize}{13}
\begin{document}
\pagestyle{fancy}
\lsstyle

\fancyhf{}
\renewcommand{\headrulewidth}{0.4pt}
\renewcommand{\footrulewidth}{0.4pt}
\setlength{\headsep}{4mm}
\fancyhead[L]{\small COOL Research DAO}
\fancyhead[R]{\small Whitepaper \monthyeardate{\today}}
\fancyfoot[R]{\small Page \thepage/\pageref*{LastPage}}

\pagenumbering{arabic}
\hyphenation{kruijs-sen}
 
\setlength{\tabcolsep}{14pt}%
\setcounter{page}{1}

\hypersetup{urlcolor=black}

\thispagestyle{empty}
\null
\vskip30mm
\begin{center}
\begin{minipage}{160mm}
\begin{center}
\includegraphics[width=\hsize]{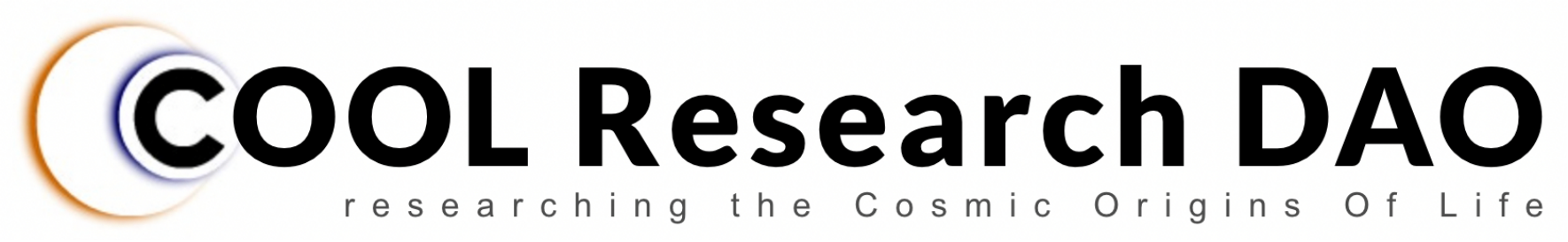}
\vskip20mm
\hrule
\vskip20mm
\fontsize{24}{24}\textbf{Whitepaper}
\vskip1cm
\fontsize{16}{16}\textbf{\monthyeardate{\today}}
\vskip10cm
\fontsize{16}{16}{M\'elanie Chevance, J.~M.~Diederik Kruijssen, Steven N.~Longmore}
\vskip0.2cm
\fontsize{16}{16}{Cosmic Origins Of Life (COOL) Research DAO
\vskip0.2cm
\href{https://coolresearch.io}{https://coolresearch.io}}
\end{center}
\end{minipage}
\end{center}
\pdfbookmark[1]{Title}{bookmark:title}


\setcounter{tocdepth}{2}
\setcounter{secnumdepth}{4}
\clearpage
\thispagestyle{empty}
\pdfbookmark[1]{Contents}{bookmark:contents}
\begingroup
\let\cleardoublepage\clearpage
\hypersetup{linkcolor=black}
\tableofcontents
\endgroup
\thispagestyle{empty}

\clearpage

\setcounter{page}{1}

\begin{center}
    {\fontsize{24}{24}\textsc{\textbf{COOL Research DAO Whitepaper}}}\vspace{1mm}\\
    {\fontsize{14.8}{14.8}\textbf{Towards community-owned astrophysics for everyone}}
\end{center}

\fontsize{\fsize}{\skipsize}\selectfont

\begin{center}
\section*{Abstract}    
\end{center}\vspace{-1mm}
\textit{Astrophysics forms a cornerstone of human curiosity and has revolutionised our understanding of the Universe. However, conventional academic structures often hinder collaboration, transparency, and discovery. We present COOL Research DAO, a next-generation framework for cosmic origins astrophysics research that uses decentralised autonomous organisation (DAO) principles to make astrophysics research globally accessible, economically fair, and intellectually open. COOL Research DAO is designed to address four key systemic problems and their manifestations in traditional academic research: general inaccessibility, hierarchical organisation, unfair economic structures, and inefficient knowledge transfer. We describe how these inefficiencies can be addressed using blockchain-based tools and token systems. By fostering bottom-up governance, open-access knowledge networks, and transparent rewards, COOL Research DAO reshapes economic, organisational, and educational paradigms in astrophysics research. Our approach leverages the data-rich and inherently international nature of astrophysics, ensuring a scalable, collaborative platform that welcomes contributions ranging from computational analyses to public outreach. In doing so, we highlight a path toward realising our vision of a truly open research ecosystem, driven by community ownership, intellectual freedom, and shared fascination with our cosmic origins.}\vspace{5mm}

\section{What problem are we solving?}
\label{sec:problem}
\pdfbookmark[1]{What problem are we solving?}{bookmark:problem}

Scientific research is the engine for discovery. Science drives humanity towards satisfying our innate curiosity, answering existential questions, and overcoming the challenges we face. While the scientific method remains the key to success, the way in which science is conducted and organised has been largely unchanged since the scientific revolution many centuries ago. This makes scientific research inefficient and sometimes even dysfunctional within the context provided by the modern world.

\begin{figure*}[tb]\vspace{-1.5mm}
\includegraphics[width=\textwidth]{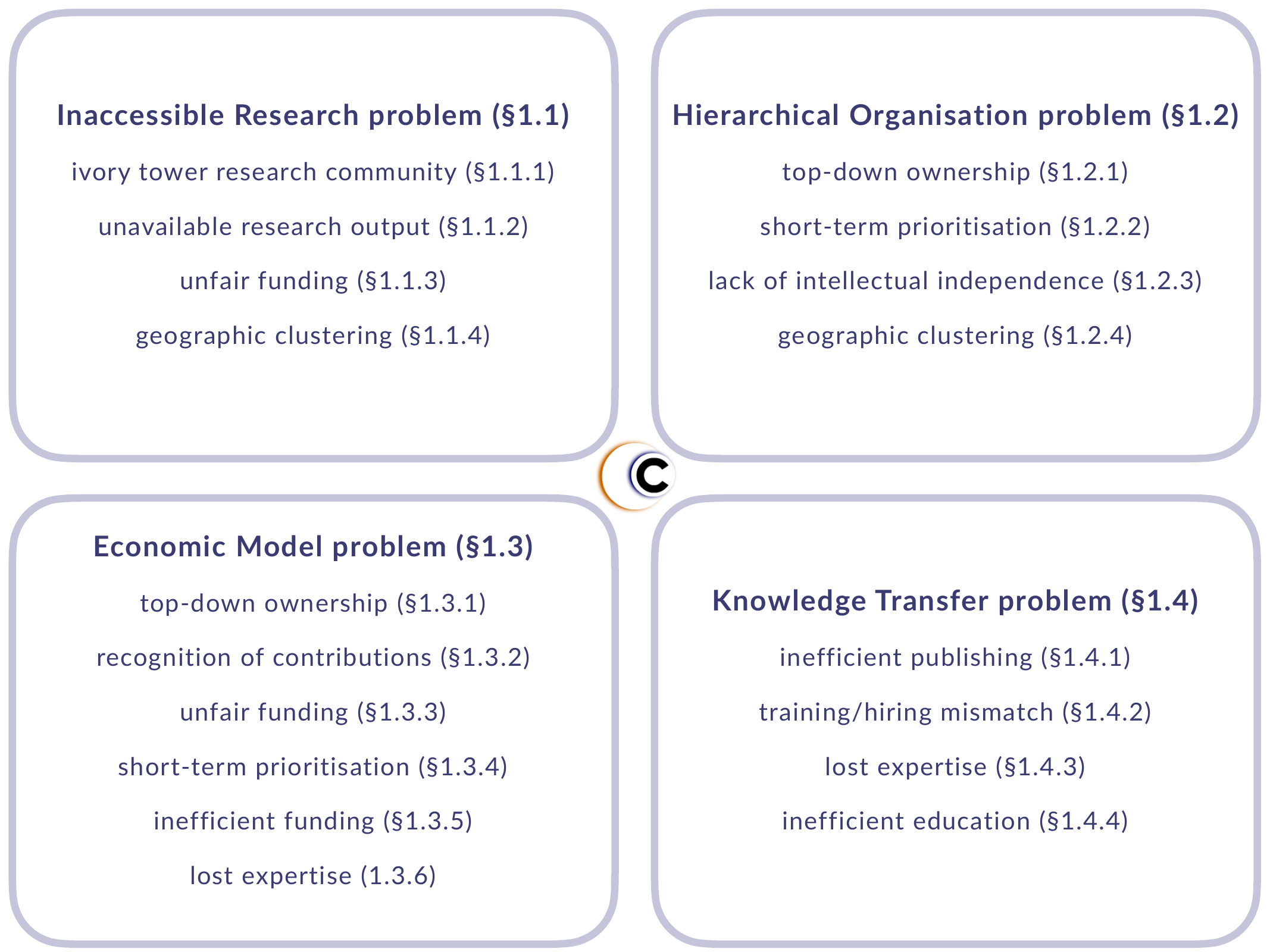}\vspace{-2mm}
\caption{\label{fig:problems}\footnotesize \textbf{Figure 1:} Summary of the four categories of fundamental problems that affect current scientific research. This whitepaper presents the \cool perspective on how these challenges can be overcome by building on recent technological developments.}
\end{figure*}

Figure~\ref{fig:problems} identifies four fundamental problems with the way in which scientific research is currently conducted and organised. Research products, communities, and funding are not commonly accessible to everyone. Research is organised hierarchically from the top down, with negative implications for funding, intellectual independence, and accessibility. Research contributions are not credited and rewarded fairly, reliably, efficiently, and sustainably. Research products and knowledge are not disseminated openly and efficiently, and the talent present in the community is not enabled to reach its full potential. Taken together, this social and organisational structure is obstructive to curiosity and discovery.

\begin{tcolorbox}
\begin{center}
\textit{Cosmic Origins Of Life (COOL) Research DAO aims to address these fundamental problems.}
\end{center}
\end{tcolorbox}

The problems faced by scientific research manifest themselves in various concrete ways within modern academia. In this Whitepaper, we provide an overview of these problems (\S\ref{sec:problem1}--\ref{sec:problem4}), present the solutions offered by COOL Research DAO (\S\ref{sec:solution}), and expand on these solutions in detail (\S\ref{sec:concrete}).

\subsection{Inaccessible Research problem} \label{sec:problem1}
Research-related activities, output, and dissemination are not openly accessible to the full range of social, economic, and geographic environments, obstructing the participation and contributions of interested and qualified potential members of the community.

\setenumerate[1]{label=\thesubsection.\arabic*}
\begin{enumerate}[leftmargin=40pt]
    \item Academic research is an ivory tower. Despite the fact that it is mostly funded by tax payers, the research community is inaccessible to a broad audience, as researchers are more likely to interact with other academics than with enthusiasts outside of academia.
    \item Publicly-funded research output is not universally openly available. This leads to data inaccessibility, irreproducibility, and a lack of societal engagement.
    \item The distribution of research funding and resources is fundamentally unfair. Investment in research and the availability of research facilities varies widely across the world, with a strong dependence on the local economic conditions.
    \item Research expertise is geographically clustered. The specialised nature of cutting-edge research results in hubs excelling in specific specialisations, restricting the personal freedom of movement of researchers and the accessibility of expertise from students or community enthusiasts.
\end{enumerate}
\setenumerate[1]{label=\arabic*.}
Solving this problem would require a global research-oriented community that is openly accessible, irrespective of the social, economic, or geographic background of interested participants.

\subsection{Hierarchical Organisation problem} \label{sec:problem2}

Modern research is not organised bottom-up around human creativity and discovery, but is shaped from the top down and is often directed by non-scientific interests, which negatively impacts intellectual priorities, independence, and ownership.

\setenumerate[1]{label=\thesubsection.\arabic*}
\begin{enumerate}[leftmargin=40pt]
    \item There is a fundamental ownership problem in academia. Research (output) is funded, governed, and owned by funding agencies, centralised research institutions, and publishers, and is often directed by large collaborations, whereas individuals have little ownership over the content that they produce.
    \item The research funding system is inflexible and unreliable. The funding schemes are organised top-down and their formats dictate the organisation of research projects. The funding is highly intermittent, enforcing the prioritisation of short-term goals and unoriginal science over visionary and potentially transformational, yet high-risk research that may take longer to realise.
    \item Research can become highly politicised. The funding and governance of modern academia permit a strong political and ideological influence, which can interfere with the intellectual independence of academic researchers.
    \item Research expertise is geographically clustered. The specialised nature of cutting-edge research results in hubs excelling in specific specialisations, restricting the personal freedom of movement of researchers and the accessibility of expertise from students or community enthusiasts.
\end{enumerate}
\setenumerate[1]{label=\arabic*.}
Solving this problem would require a decentralised, bottom-up organisation that facilitates free-form interaction and collaboration between members and allows them to focus on creativity and discovery.

\subsection{Economic Model problem} \label{sec:problem3}

The economic model adopted by the modern academic system is incapable of adequately defining scientific value, recognising contributions, and directing the funding towards scientific discovery and dissemination. Additionally, the system exhibits major inefficiencies that cause a persistent drain of value and expertise from the ecosystem.

\setenumerate[1]{label=\thesubsection.\arabic*}
\begin{enumerate}[leftmargin=40pt]
    \item There is a fundamental ownership problem in academia. Research (output) is funded, governed, and owned by funding agencies, centralised research institutions, and publishers, and is often directed by large collaborations, whereas individuals have little ownership over the content that they produce.
    \item Contributions to scientific progress are not appropriately recognised. The current system relies mostly on an inadequate and variable combination of authorship (order), citations, and conference presentations, whereas individual contributions can take many forms in the modern research ecosystem.
    \item The distribution of research funding and resources is fundamentally unfair. Investment in research and the availability of research facilities varies widely across the world, with a strong dependence on the local economic conditions.
    \item The research funding system is inflexible and unreliable. The funding schemes are organised top-down and their formats dictate the organisation of research projects. The funding is highly intermittent, enforcing the prioritisation of short-term goals and unoriginal science over visionary and potentially transformational, yet high-risk research that may take longer to realise.
    \item The research funding system is highly inefficient. The combination of short funding periods and low success rates means that considerable time and resources are lost to the preparation of funding applications.
    \item Researchers leaving academia are lost from the community. Up to 95\% of all PhDs continue their careers outside of academic research. Due to the closed nature of the academic ecosystem their expertise and involvement are lost.
\end{enumerate}
\setenumerate[1]{label=\arabic*.}
Solving this problem would require an openly accessible and diversified economic system built from the bottom up that adequately incentivises, recognises, and rewards contributions to scientific discovery and dissemination, and engages the entire community irrespective of their role or contributions.

\subsection{Knowledge Transfer problem} \label{sec:problem4}

The transfer and dissemination of knowledge in the modern academic system is highly inefficient, by adopting outdated publication formats, promoting a widespread duplication of effort, enforcing a mismatch between the training and skill sets of members and their actual (future) roles, and failing to retain built-up expertise.

\setenumerate[1]{label=\thesubsection.\arabic*}
\begin{enumerate}[leftmargin=40pt]
    \item The traditional research publication and dissemination system is highly inefficient and expensive. Journals require each publication to be entirely self-contained, whereas publications generally form a network with significant overlap in terms of context, source data, and methodology. The monolithic nature of the current publication system results in an unnecessarily slow rate of publication, as well as the duplication of information between publications. Additionally, there is little incentive for publishing null results, which can lead to considerable duplication of research efforts.
    \item There is a training-hiring mismatch in academia. Members of the academic community are trained to be world-class researchers, but most careers progress towards an increasing (and eventually dominant) commitment to teaching, funding applications, administrative tasks, and management.
    \item Higher education is usually carried out by untrained educators within a gated system. Lectures are taught by academics who often have not received any didactic training and who are not always experts in the topics that they are required to teach. This is accompanied by a redundancy in which many duplicates of individual lectures are being given to small audiences of students. At the same time, these courses have major barriers to entry in the form of language, geography, economic circumstances, or by requiring membership of the academic ecosystem.
    \item Researchers leaving academia are lost from the community. Up to 95\% of all PhDs continue their careers outside of academic research. Due to the closed nature of the academic ecosystem their expertise and involvement are lost.
\end{enumerate}
\setenumerate[1]{label=\arabic*.}
Solving this problem would require a decentralised infrastructure for free-form, peer-to-peer collaboration, education, and the dissemination of knowledge, facilitating the connection, participation, and engagement of experts in their respective disciplines, independently of their professional status or background.

\section{The COOL Research DAO solution}
\label{sec:solution}
\pdfbookmark[1]{The COOL Research DAO solution}{bookmark:solution}

\subsection{Modern technology and astrophysics: a natural frontier for solving these problems}
\label{sec:astrophysics}
Open structures on the internet are the 21st-century solution to the challenges identified in \S\ref{sec:problem}. The formation of online networks has allowed communities to self-organise, specifically in the form of decentralised autonomous organisations (DAOs).

\begin{tcolorbox}
\begin{center}
\textit{Decentralised Autonomous Organisations (DAOs) are organisational structures that make use of blockchain technology to enable their members to make consensus decisions on how the organisation is structured, how its activities can be facilitated, and how its resources are allocated. In a DAO, control is spread out across its members, instead of being built on a top-down hierarchy.}
\end{center}
\end{tcolorbox}

DAOs can solve the challenges identified in \S\ref{sec:problem} by providing decentralised, transparent, and democratic platforms for conducting and organising scientific research. With a DAO, research products, communities, and funding can be made accessible to everyone through open access and transparent allocation of resources. The hierarchical structure of research can be replaced with a decentralised governance model, allowing for more intellectual independence and accessibility to research contributions. The use of blockchain technology can enable fair, reliable, efficient, and sustainable rewards for research contributions, as well as open and efficient dissemination of research products and knowledge. Additionally, the talent present in the research community can be better utilised and retained through the collaborative and inclusive nature of a DAO. Overall, a DAO can improve the efficiency and effectiveness of scientific research, fostering a more dynamic and innovative environment for discovery.

The field of decentralised science (DeSci) is currently pioneering the use of DAOs to enable a rethinking of how science is pursued and knowledge is being disseminated. The goal of DeSci is to provide increased transparency and immutability in the tracking of funding and research progress, allow for greater collaboration and data sharing among community members, as well as increased transparency and security for the management of scientific data. Within a DAO, it is possible to form topical communities (instead of communities organised by the level in the social hierarchy) connecting e.g.\ researchers, public outreach experts, (audio-visual) artists, content creators, citizen scientists, investors, and enthusiasts to disseminate knowledge and share intellectual freedom and discovery within their topical area(s) of interest.

Given the differences in research culture, resources, and infrastructure between different subject areas of modern research, some areas are more naturally suited for the translation to DeSci than others. The Cosmic Origins Of Life (COOL) Research DAO represents a case study of these solutions in a topical area that is ideally suited for pushing the boundaries of the scientific infrastructure. The astrophysics of our cosmic origins forms a natural environment for pioneering the push towards an open, global research community. A non-exhaustive set of reasons is as follows.
\begin{enumerate}
    \item \textit{Universal appeal:} The questions surrounding our cosmic origins are existential, universally compelling, and transcend geographical, cultural, and social boundaries. This common curiosity creates a strong foundation for a global research community.
    \item \textit{Non-territorial research area:} Because the subject of cosmic origins astrophysics lies beyond geopolitical boundaries, it symbolises a global endeavour and is fitting for a platform that similarly knows no borders.
    \item \textit{Public engagement:} The field is known for its ability to engage public interest, which can help both to popularise the concept of a DAO and to potentially crowdsource research or data analysis tasks.
    \item \textit{Interdisciplinary nature:} The field inherently involves a variety of scientific disciplines -- physics, chemistry, geology, biology, computer science, and even philosophy. This naturally encourages a collaborative, global approach and a DAO can facilitate this interaction.
    \item \textit{Complexity of research:} The complexity and scale of the problems being addressed in astrophysics (such as the origins and evolution of life in the Universe) require collaboration and the efficient exchange of ideas on a global scale.
    \item \textit{Long-term research horizons:} Astrophysics involves long timescales, both in terms of the phenomena being studied and the timeline of research projects. This fits well with a DAO, which benefits from long-term stability and planning.
    \item \textit{Dependence on large infrastructure:} Much of astrophysics research relies on large, often multi-national projects (e.g.\ major space telescopes). A DAO can build upon and further strengthen this established culture of international collaboration.
    \item \textit{Reliance on remote-access facilities:} Astrophysics relies heavily on remote access to major observational or supercomputing facilities, making it well-suited to a decentralised global community. Researchers from around the world can collaboratively utilise these facilities, enhancing the efficiency and quality of the research output.
    \item \textit{Data abundance:} Astrophysics is a data-rich field. The rapid increase of data volume, production rate, and complexity requires global cooperation and shared resources to process and interpret these data.
    \item \textit{Open access policies:} Many astrophysical data sources and software packages are publicly available. This aligns well with the philosophy of an open research community.
\end{enumerate}
Given these considerations, \cool is ideally positioned to help spearhead the transformation of academia towards a decentralised research-oriented ecosystem, representing a natural test bed for these innovations.

\subsection{Envisioning a decentralised research-oriented ecosystem}
The \cool ecosystem will enable a transformative shift of the traditional academic system towards an open, research-oriented community. This is achieved by establishing a decentralised autonomous organisation (DAO) that addresses the persistent challenges of the traditional system with innovative solutions, while retaining the current infrastructure of essential physical hubs (e.g.\ laboratories, instrumentation, computing centres). \cool aims to reshape intellectual, economic, and educational paradigms in academia, using blockchain technology as a foundation for transparency, inclusivity, and efficiency.

Incorporating and acknowledging contributions outside the rigid bounds of the current academic hierarchy, \cool plans to redefine the existing academic economic model, promoting diversified funding avenues, providing fair ownership, recognition, and rewards. Through this new infrastructure, the system will move away from the PI-driven model, fostering intellectual independence and entrepreneurship, facilitating granular attribution, and encouraging individual agency. \cool intends to streamline knowledge transfer by establishing an open standard for knowledge exchange and deploying a global, decentralised network for education that optimises the alignment of tasks, skills, and interests.

Ultimately, \cool seeks to move academic research beyond its traditional confines and cultivate a more equitable, productive, and enriching academic culture. The following sections detail the various strategies \cool proposes to tackle the innate problems of the traditional academic system.

\section{Concrete Implementation of the COOL Research DAO Solution}
\label{sec:concrete}

\subsection{Addressing the Inaccessible Research problem}
The problem laid out in \S\ref{sec:problem1} can be solved by building a global research-oriented community in the form of a (network of) DAO(s) that is openly accessible, irrespective of the social, educational, economic, or geographic background of potential participants. Such an organisation enables better, more open communication, it maximises the progress made within its topical area by allowing and rewarding contributions from non-academic participants, and helps facilitate a fair distribution of funding and resources.

\paragraph*{An accessible community} Once the full governance structure is in place, participation in \cool will be open to researchers, public outreach experts, content creators, citizen scientists, and enthusiasts, irrespective of their background or education. Communication between members is facilitated by an online platform that enables written and spoken interaction. At \cool, this role is initially fulfilled by a Discord server on which members are encouraged to engage with one another and seek feedback from other members throughout the lifetimes of the projects that they are working on. It may be enriched with other infrastructure, such as a \cool `Hub' on platforms such as \textit{ResearchHub} (\href{https://www.researchhub.com}{researchhub.com}). The online infrastructure helps facilitate collaboration and co-creation between researchers from different regions and countries, and supports the sharing of expertise and knowledge across geographical boundaries. Contributions can take many forms, such as providing ideas, carrying out research, generating scientific or public outreach content, creating courses, giving lectures, or organising discussions. These contributions are rewarded in the form of tradable tokens. Additionally, tokens can fulfil a governance role, enabling members to participate in the governance process and achieving transparency in decision making, project selection, and funding. 

\paragraph*{An accessible, equitable, and transparent funding system} As discussed in \S\ref{sec:problem1}, the current research funding system needs an overhaul to accommodate new ways of evaluating and recognising the value of research contributions, which are more comprehensive and inclusive than traditional metrics such as past funding, citations, and journal impact factors. The decentralised ecosystem of \cool makes funding more accessible by allowing funding to be exchanged democratically based on the merit of the idea, the project, and the team, and provides the infrastructure for each project to distribute the resources fairly to all participants. In practice, this implies a mechanism and platform for the allocation of research funding and resources that is more transparent and equitable, facilitating crowdsourcing, crowdfunding, angel investments, and token-based reward systems. These mechanisms help to democratise the process of funding research, and at the same time make it more responsive to the needs and interests of researchers and other stakeholders, including those from under-resourced regions and countries. That way, it also breaks away from the boundaries imposed by the currently dominant role of local funding bodies, which is especially important in view of the increasingly international compositions of project teams. Existing platforms like \textit{ResearchHub} could be used to help fulfil this role, but these typically target participation by researchers. Achieving all of the above goals will require involving the broader community (specifically to enable crowdfunding and angel investments), which will either require the expansion of such platforms or the integration of additional ones.

\paragraph*{Permanent and openly-accessible project outputs} Finally, \cool will adopt open-access policies and publication practices to overcome the inaccessibility of research output. The solution to this problem is not to introduce high open-access charges at siloed journals, but to move the entire publication and refereeing system to a transparent, peer-to-peer, and open access platform. At \cool, this platform should eventually be provided by blockchain technology, which enables the permanent and transparent storage of research output at a transaction fee that represents only a negligible fraction of traditional page charges. Moreover, this medium facilitates viable deviations from traditional publication formats, such as the publication of short ideas, failed experiments, code, and data products. The network nature of this infrastructure also enables the attribution of value and recognition to research output based on the links to other nodes in the network, which is fairer, more comprehensive, and more transparent than the current recognition system based solely on published paper citations and journal impact factors.

\subsection{Addressing the Hierarchical Organisation problem}
The problem laid out in \S\ref{sec:problem2} can be solved by building a decentralised, bottom-up organisation in the form of a DAO, to facilitate free-form interaction and collaboration between members, allowing them to focus on creativity and discovery. Such an organisation enables a representative and comprehensive attribution of contributions, the more organic and effective planning of projects and collaborations, and protects the intellectual independence of its members.

\paragraph*{A decentralised organisation of research institutions and collaborations} \cool aims to adopt a network structure for the interaction among its members to stimulate bottom-up interaction and collaboration. Such forms of interaction are common in collaborations with few members, but are increasingly lacking in the academic ecosystem due to the rise of mega-collaborations and the growing influence of geographically centralised research hubs. At \cool, the decentralised network constituted by its members is fully scalable, such that interaction remains free-form and bottom-up irrespective of the total size of the network. The research agenda and strategy emerge from the bottom up, based on individual interests and needs, free from ideological or political influences. Every member has full intellectual autonomy and is free to instigate and carry out projects. By using online infrastructure, \cool aims to facilitate the self-organisation of its members purely based on their interests and initiative, irrespective of their geographical location.

\paragraph*{A decentralised organisation of research funding and support} The ecosystem of \cool enables and supports the development of new funding structures, that are driven bottom-up by the individual needs and interests of its members and their projects rather than the top-down restrictions imposed by funding bodies. This way, the funding system is more observant of and responsive to the natural timeframe, scope, and cost of each individual project. Economically, this infrastructure will be enabled by crowdsourcing, crowdfunding, angel investments, and token-based reward systems within the ecosystem (see \S\ref{sec:solution3}). The natural flexibility with which such funding sources can be allocated and used makes them highly scalable as the size of a collaboration evolves and the project is carried out. In such a system, the probability that a project is funded becomes proportional to the ability of a project to convey its scientific relevance and engage the broader (non-academic) community. This also presents a major opportunity for fundamental research with a longer time horizon, because the bespoke project timeframes within the \cool ecosystem enable the funding and realisation of high-risk projects, contrary to any current funding scheme provided by traditional funding bodies.

\paragraph*{Novel ways of evaluating and recognising contributions} The online infrastructure underpinning \cool is well-suited for adopting more organic forms of tracking, evaluating, and recognising contributions, going beyond the traditional attribution system that is imposed and propagated by the top-down organisation of current academic research. Currently, project outputs are treated as a single monolithic product, which is then largely attributed to the principal investigator (PI) of a research project. The transparent and more granular attribution of contributions to individual members of a collaboration would represent a major improvement relative to the current dominant role of the PI. Within the ecosystem of \cool, it will be possible to attribute e.g.\ participation in discussions, contributions to ideas, the realisation of null results, the replication of new results, contributions from outside the research community, public outreach, and many other contributions that often represent major achievements but currently cannot be adequately attributed to individuals. The use of blockchain technology would then additionally enable mapping the network generated by the impact of these individual contributions throughout the topical domain of \cool, which facilitates the conversion of fair ownership to the fair crediting of contributions (see \S\ref{sec:solution3}). Finally, foregoing the PI-driven model protects the intellectual independence of participants across the community and provides agency to individuals to shape their own careers. This way, the \cool ecosystem promotes interaction-driven, bottom-up creativity between its members by enabling intellectual entrepreneurship, fair ownership, and bespoke project timeframes.

\subsection{Addressing the Economic Model problem} \label{sec:solution3}
The problem laid out in \S\ref{sec:problem3} can be solved by building an openly accessible and diversified economic system in the form of a DAO, built from the bottom up, that adequately incentivises, recognises, and rewards contributions to scientific discovery and dissemination, and engages the entire community irrespective of their role or contributions. This economic system acknowledges and utilises the intrinsic value of knowledge and content. It admits that modern academia already operates as a competitive market and addresses the fundamental inefficiencies of this market by allowing its members to be free actors, creating new stores of value, and generating the functional liquidity for a highly efficient, flexible, and sustainable ecosystem.

\paragraph*{A highly efficient, inclusive, and decentralised economic model} 
The impact, utility, and economic value of a network scale with the number of participants squared. The \cool ecosystem aims to capitalise on this empirical law by adopting a decentralised, granular economic infrastructure to facilitate a continuous exchange of value through the network. It will address the fundamental inefficiency of the centralised, monolithic academic ecosystem by overcoming its intrinsic inertia and illiquidity. By building an on-chain, decentralised ecosystem, \cool aims to adopt peer-to-peer infrastructure for the exchange of intellectual value through tokenisation. This infrastructure enables a high rate of transactions of any size, thereby generating a new intellectual economy within which value can be exchanged continuously rather than in top-down imposed quanta, that can adapt more quickly to the needs of individual actors, that can facilitate a fairer distribution of resources, that can support projects from the bottom up regardless of their size, and that can bring resources of any kind and from anyone directly into the economy without relying on top-down governance structures. Such an economy is highly inclusive by definition, and allows participants to remain in the network regardless of their role within the ecosystem (e.g.\ academic or non-academic), thus retaining their value even when they change roles (e.g. when they move from academia to industry).

\paragraph*{A transparent economic model through fair attribution of credit}
The decentralised, granular, and continuous economy of the \cool ecosystem aims to enable the transparent, fair, and unambiguous attribution of credit, thereby unlocking a wide array of forms of compensation and recognition. Fundamentally, each action within the network can be treated as a transaction; when a member contributes to a project and adds value, that member receives a store of value in return. This can be in the form of a non-fungible token specifying the nature of the contribution, which could also serve as a record of intellectual property, or a fungible token that can be used for any other purpose. Each of these rewards unlock a near-infinite number of applications. A non-fungible token specifying the contribution made can be used as an access pass or invitation to future events, lectures, or other projects. It can be used as a key for receiving rewards whenever the project results make a contribution to other projects. A member's collection of non-fungible contributions automatically forms a record that can be used to evaluate their profile when they apply for joining projects or seek funding. These perks persist even when a member changes roles. Conversely, a fungible token reward can be used to fund other projects, it can be exchanged for non-fungible tokens, goods, or services within the ecosystem, and it can be converted to real-world currency. Many of the above use cases exist in some form in the current academic system (e.g.\ recruitment, funding, reward structures, conference invitations, citations, curriculum vitae), but the above realisation of such mechanics within the \cool ecosystem would represent a far-reaching improvement in terms of transparency, fairness, inclusivity, intellectual freedom, attribution, and recognition.

\paragraph*{New forms of funding and support for research}
The open and decentralised economy envisioned for \cool will offer multiple benefits relative to the current academic funding system. First, the existing funding infrastructure for research and dissemination can naturally be integrated into the \cool ecosystem. On the one hand, the infrastructure is well-supported and thus represents one of several sources of funding to the ecosystem. On the other hand, the decentralised infrastructure of \cool and its focus on non-hierarchical accessibility, fair recognition, and transparent ownership can serve as the ideal environment for traditional funding bodies to inform their decision making and funding allocation from the bottom up. It can also offer immediate exposure of these funding bodies and organisations to the global intellectual market, thus reducing expenses and increasing efficiency. Secondly, the open and decentralised nature of \cool has the potential to unlock the integration of a wide variety of novel funding streams within the same ecosystem. By allowing transactions of any size, the infrastructure would naturally enable the direct incorporation of crowdfunding, angel investments, charities, as well as media and industry partnerships into the global research economy. This would enable a seamless combination of centralised and decentralised funding sources that is more responsive and accountable to the needs and interests of individual researchers and projects. It would support the realisation of visionary projects by connecting to funding providers with long-term conviction. It would enable the extension of the academic research economy into dissemination, education, public outreach, content creation, popular culture, and industry. It will allow geographically restricted communities built around local research and public outreach projects to participate in the global research economy from anywhere in the world. This way, the \cool infrastructure will contribute to the fair and bespoke distribution of funding, attuned to the individual needs and time horizon of each project. These solutions would provide the ingredients for a highly efficient, flexible and sustainable ecosystem that is fundamentally enabled by the extreme liquidity of a decentralised global research economy.

\subsection{Addressing the Knowledge Transfer problem}
The problem laid out in \S\ref{sec:problem4} can be solved by building a decentralised infrastructure in the form of a DAO to support free-form, peer-to-peer collaboration, education, and the dissemination of knowledge, which facilitates the connection, participation, and engagement of experts in their respective disciplines, independently of their professional status or background.

\paragraph*{An efficient network for the dissemination of knowledge, discovery and ideas}
\cool strives for the construction of a new open standard for the exchange of knowledge that overcomes the major shortcomings of the current academic publishing system. \cool aims to adopt an openly-accessible, transparent, and granular network of research contributions, ideas, and results of any size, which can be linked to any other set of contributions, and to which anyone can contribute. This infrastructure needs to enable the publication and recognition of individual results (and their replication), data sets, contributions to discussions, scientific images, null results, and any other form of contribution, without the formal requirements imposed by the current scientific publication system. Each contribution within the network can be connected to other contributions manually or automatically, thereby providing topical and methodological context that is commonly duplicated in traditional publications. Within this new publishing infrastructure, the traditional form of peer review is superseded by a more organic and transparent process, in which follow-on discussions form an integral part of the ecosystem and more valuable contributions are used more often. Together, these feedback loops represent a form of community review of a contribution, which is fairer, more efficient, and more constructive than the single point of failure constituted by traditional peer review performed by a small number of experts. 

By placing the contributions on-chain, it will be possible to construct a transparent network of contributions, in which each node connects to previous and subsequent results. Such a structure would enable the fair recognition and crediting of contributions. To navigate this increasingly complex network of knowledge, artificial intelligence (AI) agents can help organise contributions and act as a navigation assistant, tailored to the individual needs of each participant. It can instantaneously summarise the relevant parts of the network of contributions to generate context-specific introductory material. Traditional publications commonly duplicate context or provide a biased perspective. Instead, AI-generated context provides a balanced perspective, prevents duplication, and allows contributions in the network to focus solely on their unique added value. This will help close the gap between state-of-the-art research and the global community of astrophysicists, data scientists, citizen scientists, public outreach experts, content creators, artists, and other enthusiasts.

Altogether, the resulting knowledge network of \cool is fundamentally open-access, optimises the efficiency of dissemination, facilitates the rapid exchange of results, minimises overlap and duplication, leverages community review, enables continuous feedback, facilitates the transparent tracking and crediting of contributions, retains author ownership, breaks free from the top-down monopoly of publishers, allows funding providers to be more precisely connected to individual contributions, and even allows public outreach experts to efficiently track and disseminate contributions outside of the traditional scientific community through popular articles, art, culture, or any other form of translation. This comprehensive vision of open knowledge exchange that has long been held as the gold standard by academic researchers and is now finally within reach.

\paragraph*{Optimised matching of tasks, skills, and interests} 
Across the world, decentralised communities have already been able to self-organise and optimise the task distribution according to the skill sets of its members. The skill-task mismatch in the traditional academic system can naturally be overcome by leveraging the global decentralised network of \cool, as it connects members with a wide variety of complementary roles and skill sets. This enables tasks to be matched to members with the necessary skills and interests. Each task contributes towards the common goal of furthering knowledge across the network. The economic infrastructure envisioned by \cool will naturally incentivise and reward tasks like teaching, management, and administrative work, alongside traditional research. Across the community, this will enable each member to focus on the activities in which they are the most interested and proficient, for the duration that best matches their needs. The reach of \cool beyond the traditional academic domain uniquely enables non-academic experts in key operational activities to make important contributions to the accumulation of knowledge through research, discovery, and dissemination. Anyone will be able to organise conferences, workshops, and lectures, set up administrative infrastructure, participate in the preparation of funding requests, or contribute to community management. The optimised assignment of tasks will allow \cool members interested in research to dedicate a larger part of their time to research activities than in the traditional academic system. In addition, researchers who would leave traditional academia will not be lost from the community, but remain an integral part of the \cool ecosystem. By continuing to seamlessly incorporate their contributions, \cool alleviates the inefficient cycle of losing valuable expertise and subsequently retraining that continuously hampers progress in traditional academia.

\paragraph*{An improved higher education system}
Successful education requires combining cutting-edge domain expertise in the subject matter with the proficient transfer of that knowledge, while optimising the reach among its recipients. The network infrastructure of \cool is aimed at connecting world-leading researchers to trained educators and the global pool of interested learners. \cool will use a decentralised, open-access, online education platform, where knowledge can be shared broadly, encouraging peer-to-peer collaboration and learning. Topical experts without the time or inclination to teach will be able to easily identify and collaborate with trained educators to create courses that are topically relevant, highly engaging, and efficient at relaying knowledge to a broad audience of participants. Experts with a natural aptitude for teaching will have an unprecedented opportunity to share their enthusiasm, while trained educators soliciting topical domain knowledge will be able to access a large pool of experts. Conversely, prospective students will have access to a wide variety of courses, allowing them to choose the ones that best match their interests, methods of learning, and time commitment, while also having the opportunity to engage with the latest advances in the field. The optimised access between experts, educators, and participants minimises redundancy between courses, overcoming a major inefficiency of the traditional higher education system. The global education network of \cool will be inherently versatile, enabling a wide variety of hybrid setups, such as combining online, globally accessible lectures with small-scale, in-person discussion groups and tutorials within local communities to maintain face-to-face interaction. Finally, the decentralised nature of the \cool education network would naturally overcome the main barriers to entry by maximising accessibility irrespective of language, disabilities, economic conditions, geographic location, or membership of the traditional academic system.

\section{Conclusion}
\label{sec:conclusion}
\pdfbookmark[1]{Conclusion}{bookmark:conclusion}

This whitepaper discusses long-standing barriers in traditional academic structures and proposes solutions that leverage modern technology to reimagine the scientific participation model, making scientific research truly accessible, equitable, and impactful. By building upon a decentralised autonomous organisation structure, \cool provides an alternative to top-down research practices, fragmentary collaboration, and the hierarchies that often hinder creativity, ownership, and credit. We envision a global, open-access research community where membership transcends economic, social, or geographical limitations, and where contributions of all forms (ranging from e.g.\ computational data analysis to public outreach) are formally acknowledged and rewarded.

A key goal of \cool is to reinvent the economics of science with a concrete focus on astrophysics, where large-scale data and infrastructure often require long-term, international collaboration. Token-based reward systems, granular attribution of contributions, and transparent funding allocations allow both traditional stakeholders (universities, agencies, and large observatories) and emerging groups (citizen scientists, private foundations, and industry) to converge around a shared system that prioritises fair value exchange and intellectual independence. This new research economy is inherently scalable, enabling projects to secure bespoke funding paths that reflect their true complexity or timescale, irrespective of whether this is a short, data-focused sprint or a decades-long endeavour to explore fundamental questions about our cosmic origins.

Furthermore, \cool rethinks knowledge exchange. Instead of siloed publication practices and opaque review procedures, the network structure provides open channels for discovery, discussion, and replication. Each idea or result, no matter how small, can be transparently recorded, peer-evaluated, and linked to the broader research ecosystem. This not only mitigates duplication of effort, but also acknowledges the cumulative value of incremental progress, failed trials, and interdisciplinary crossovers. Next to a wide variety of online education tools and global peer-to-peer mentorship, \cool envisions a vertically-integrated ecosystem where anyone (e.g.\ researchers, public outreach experts, artists, content creators, investors, citizen scientists, or enthusiasts) can directly contribute to and benefit from the continuous flow of new insights.

By combining these innovations with the deep human fascination for the cosmos, \cool offers a natural proving ground for decentralised science. Astrophysics already thrives on extensive international cooperation and data sharing. The \cool approach simply provides new structural and economic frameworks to make this cooperation seamless, inclusive, and long-lived. The end result is a robust and modern ecosystem in which the quest for knowledge, from cosmology to the search for life beyond Earth, becomes a community-owned endeavour. \cool aims to transform astrophysics into an open, thriving enterprise where curiosity and collaboration transcend the hierarchies and limits of geography, politics, and traditional funding models.

\end{document}